\documentclass[aps,prl,preprint,superscriptaddress]{revtex4-1}

\usepackage{graphicx}

\begin{document}


\title{Propinquity drives the emergence of network structure and density}


\author{Lazaros K. Gallos}
\affiliation{DIMACS, Rutgers University - Piscataway, NJ 08854, USA}
\author{Shlomo Havlin}
\affiliation{Department of Physics, Bar-Ilan University, Ramat Gan 52900, Israel}
\author{H. Eugene Stanley}
\affiliation{Physics Department and Center for Polymer Studies, Boston University, Boston,
Massachusetts 02215, USA}
\author{Nina H. Fefferman}
\affiliation{Departments of Mathematics \& Ecology and Evolutionary Biology, University of Tennessee, Knoxville, Tennessee 37996, USA}

\begin{abstract}
The lack of large-scale, continuously evolving empirical data usually limits the study of networks to the analysis of snapshots in time. This approach has been used for verification of network evolution mechanisms, such as preferential attachment. However, these studies
are mostly restricted to the analysis of the first links established by a new node in the network and typically ignore connections made after each node’s initial introduction. Here, we show that the subsequent actions of individuals, such as their second network link,
are not random and can be decoupled from the mechanism behind the first network link. We show that this feature has strong influence on the network topology. Moreover, snapshots in time can now provide information on the mechanism used to establish the second
connection. We interpret these empirical results by introducing the `propinquity model', in which we control and vary the distance of the second link established by a new node, and find that this can lead to networks with tunable density scaling, as found in real networks. Our work shows that sociologically meaningful mechanisms are influencing network evolution and provides indications of the importance of measuring the distance between successive connections.
\end{abstract}

\maketitle

\section*{Introduction}
The explosion in network research has been largely driven by the availability of big social data, the analysis of  social systems, and by studying the mechanisms behind the emergence of behavioral networks\cite{1,2,3,4,5,6,7,8,9,10}. Network generation methods are central in modeling network evolution and have helped us understand many properties of these systems, even when only a static snapshot is available. There exists a large variety of mechanisms which have been proposed and verified\cite{11}, such as the famous preferential attachment principle \cite{12}, where nodes connect with higher probability to higher connected nodes. Different requirements may be imposed, such as requiring an unbiased configuration \cite{van2018sparse}, and the mechanisms are usually adapted to the empirical systems that they attempt to explain.

In a typical network evolution model, new nodes are introduced into the system and they become connected to existing nodes according to certain rules. It is also possible that further changes can take place in the network, such as redirection of existing links, introduction of new links among existing nodes, etc. Recently, for example, Redner et al. \cite{13,14} studied a copying model, which is based on duplication-divergence mechanisms \cite{15}, and showed that a new node that inherits a fraction of connections from its first link can give rise to a diversity of topologies, mainly in terms of network density.

In the majority of these models, the rules for attaching a node specifically target the identification of the first connection. When a new node creates more than one connection then the same rules are typically applied to identify each one of those connections, e.g., a node connects to $m$ nodes via preferential attachment \cite{12}. However, in a real evolving system the agents continue adding links for a long time after they are introduced in the network and it is highly unlikely that the processes of initial introduction are simply replicated over the complete lifespan of a node. This process of adding additional links is probably too complicated to observe in real networks or to model accurately. However, there is a tractable important question about the distance between the first $m$ connections of a new node which has not been explicitly addressed, even though it may be a key factor in defining central network properties, such as the network density.

Here, we present a first step that considers mechanisms that influence the choice of the {\it second} connection for newly introduced nodes. We suggest a model that can quite accurately capture the behavior of real-world evolution in empirical networks. The mechanism that we introduce here restricts the {\it distance} between the first and second connections of a new node, as measured prior to the node's introduction. As we show, the resulting network topology depends on the proximity of these two connections; we therefore call this the `propinquity model'.

\begin{figure}
\centering
\includegraphics[width=.95\linewidth]{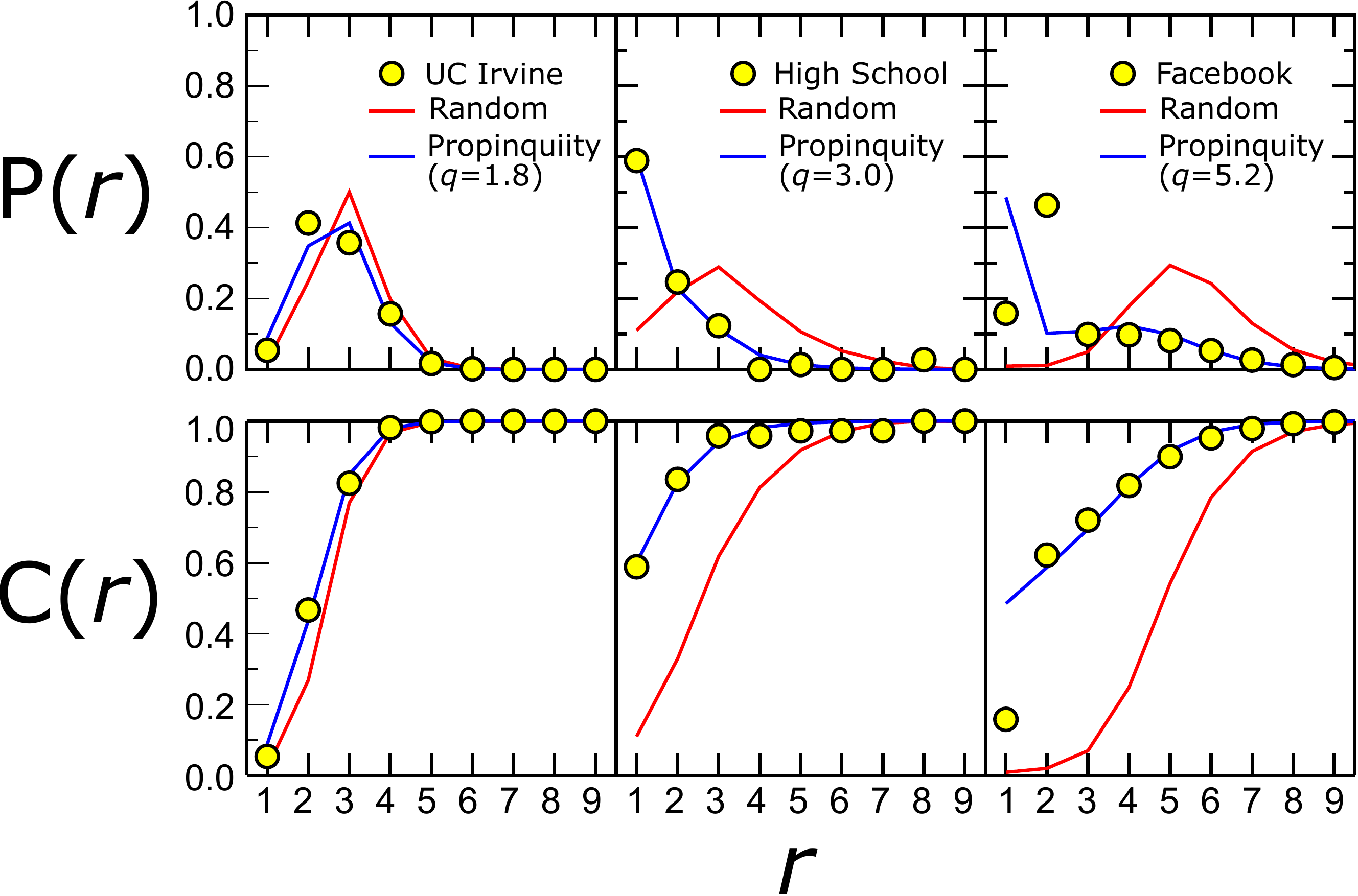}
\caption{Probability that the second link connects to a node at a distance $r$ from the first node. We measure the probability distribution (top) and the cumulative distribution (bottom) of the distance $r$ between the first two neighbors of a new node in a network. Left to right: online social network in UC Irvine\cite{16} ($N=1893$), High School friendship\cite{17} ($N=180$), and Facebook wall messages \cite{18} ($N=43953$). Symbols represent the empirical results. The red lines correspond to the case where the second node would be selected in random ($q=0$ in the propinquity model, see Fig. \protect\ref{fig2}). The exponents $q$ in the propinquity model that give the best fit to the real data are shown in blue and represent the tendency of the distance to be smaller than random. Note that propinquity does not explain well the dynamics driving the Facebook wall messaging network for $r=1$ and $r=2$, but works well for larger $r$. The origin for this could be that initiation into a wall message network may be impacted more strongly by influence external to the online network (i.e. alternative means of communicating with friends, and need for communication with friends-of-friends, apart from Facebook wall messages) and thus slightly skewing the results. Notice however that the total probability for $r\leq2$ is still consistent with the model prediction. Propinquity nevertheless offers meaningful and valuable insight as $r$ increases.}
\label{fig1}
\end{figure}

As a first demonstration that this metric can provide meaningful insight, we show that this distance does not behave trivially in empirical networks (Fig.~\ref{fig1}). The network evolution in the three presented networks is known and we are therefore able to measure the distance between the first two connections for each new node just prior to the node's introduction. The resulting distance distribution cannot be characterized by a uniform distribution within the network, i.e. the distance of the second connection is not a randomly chosen quantity. On the contrary, each network seems to have its own characteristic distribution for these distances. In social networks, for example, shorter distances seem to be significantly preferred.

Using the underlying concept of our propinquity model, in this paper we explain the observed distance distributions in Fig.~\ref{fig1} and use this insight to propose proximity as a novel metric for characterizing the ongoing social dynamics of evolving networks in meaningful behavioral ways.  We show how this characterization can lead to a systematic variation of network density, and we can use this metric to distinguish between network structures even when quantities such as the degree distribution and clustering coefficient seem identical.

\section*{Model Description}

\subsection*{Local network density}

The underlying principle of network theory is that link structure among nodes provides more information than could be learned by examination of the nodes in isolation \cite{19}. In other words, connectivity is the main factor that determines the network behavior and response. Typical methods used to estimate the organization of links include, e.g., modularity or community detection \cite{20}, fractal properties \cite{21}, transport properties \cite{22}, percolation properties \cite{23,24,25} etc. Surprisingly, little work has been done on direct measurements of link density in real networks (see e.g. the concept of $n$-tangle density \cite{26}). However, the above approaches are mainly descriptive rather than predictive and there is currently no generic framework to detect potential mechanisms that explain the variation of local densities, especially at different system scales.

In terms of characterizing emergent density properties, there are two main families of growing network models. The most common mechanisms add a constant number of links for each node and, as a result, the link density is also constant, easy to calculate, and is rarely given any further consideration (this is e.g. the case of the preferential attachment mechanism \cite{12}). The second family of models uses a probabilistic mechanism of adding new links and can lead to either sparse or dense networks depending on the model parameters (such as duplication-divergence models \cite{27}).  In contrast to these two general cases, the propinquity model leads to networks that have a known global density, but (in contrast to earlier models) simultaneously enables a systematic variation of local density at different scales, as observed in real networks.

By focusing on the time-ordered behavior of local links and the resulting local density, and how this varies at different scales within the network, we can explain the emergence of communities and understand differences in the types of social dynamics that we observe in real-world networks. To quantify this local link density, scale is determined by the number of nodes, $n$, in a connected subgraph of the network. Formally, the link density $\rho$ in a graph with $N$ nodes and $L$ links is defined as the fraction of the number of links over the maximum possible number of links \cite{28}, i.e. $\rho=L/[N(N-1)/2]$. To measure the local link density we consider an induced connected subgraph of $n$ nodes, where we take into account all the $e_n$ existing links between all pairs of nodes in the subgraph. We then define the local link density as,
\begin{equation}
    \rho_n = \frac{e_n}{\frac{n(n-1)}{2}} .
    \label{Eq1}
\end{equation}

This allows us to study scaling of local link density as we vary the size of the subgraph, $n$. As explained in detail in the SI, the behavior of this quantity is highly influenced by a trivial property. This is because we restrict ourselves to connected subgraphs of size $n$, which by definition requires all the subgraphs to have at least $n-1$ links for connectivity. The simple solution that has been suggested is to subtract $n-1$ links from the numerator in Eq. (\ref{Eq1}) \cite{26}. In real networks the density has been shown to scale inverse linearly with the network size, i.e. $\rho_n\sim n^{-1} + O(n^{-2})$ \cite{29,30}. This means that $e_n \sim n + O(1+n^{-1})$ and the linear term dominates the behavior of $e_n$. For sparse networks where the prefactor of $n$ is close to 1, if we simply subtract these links from $e_n$, the density behavior will now depend on the higher-order terms, which may scale in a different way than $\rho_n$. We therefore apply here the recently defined metric \cite{26} for the local $n$-tangle (Topological Analysis of Network subGraph Link/Edge) density, $t_n$, as
\begin{equation}
    t_n = \frac{e_n-(n-1)}{\frac{n(n-1)}{2}-(n-1)} .
    \label{Eq2}
\end{equation}
The key feature in this definition is the removal of the $n-1$ links that are necessarily present in an induced subgraph to ensure connectivity. We also remove the same number of links in the denominator, so that $t_n$ remains properly normalized and ranges from $t_n=0$ in the case of a loopless tree subgraph to $t_n=1$ for a fully connected subgraph.

\subsection*{The predictive power of the propinquity model}
For the model to be useful as a predictive tool, we must be able to validate hypotheses about the ways in which new nodes choose to connect to the network by agreement with observations of real-world network structures. There is already a large variety of network growing models in the literature \cite{12,31,32,33,34}. Typically, starting from a seed network, new nodes are introduced and attach themselves according to certain rules, e.g. by connecting preferentially to the most connected nodes. However, in many real systems nodes have a restricted freedom or ability to reach all the available connections \cite{35}, thus the ability to create meaningful, behavioral-hypothesis driven growing models would vastly expand our toolkit for understanding the mechanisms of ongoing social dynamics.

\begin{figure}
\centering
\includegraphics[width=.95\linewidth]{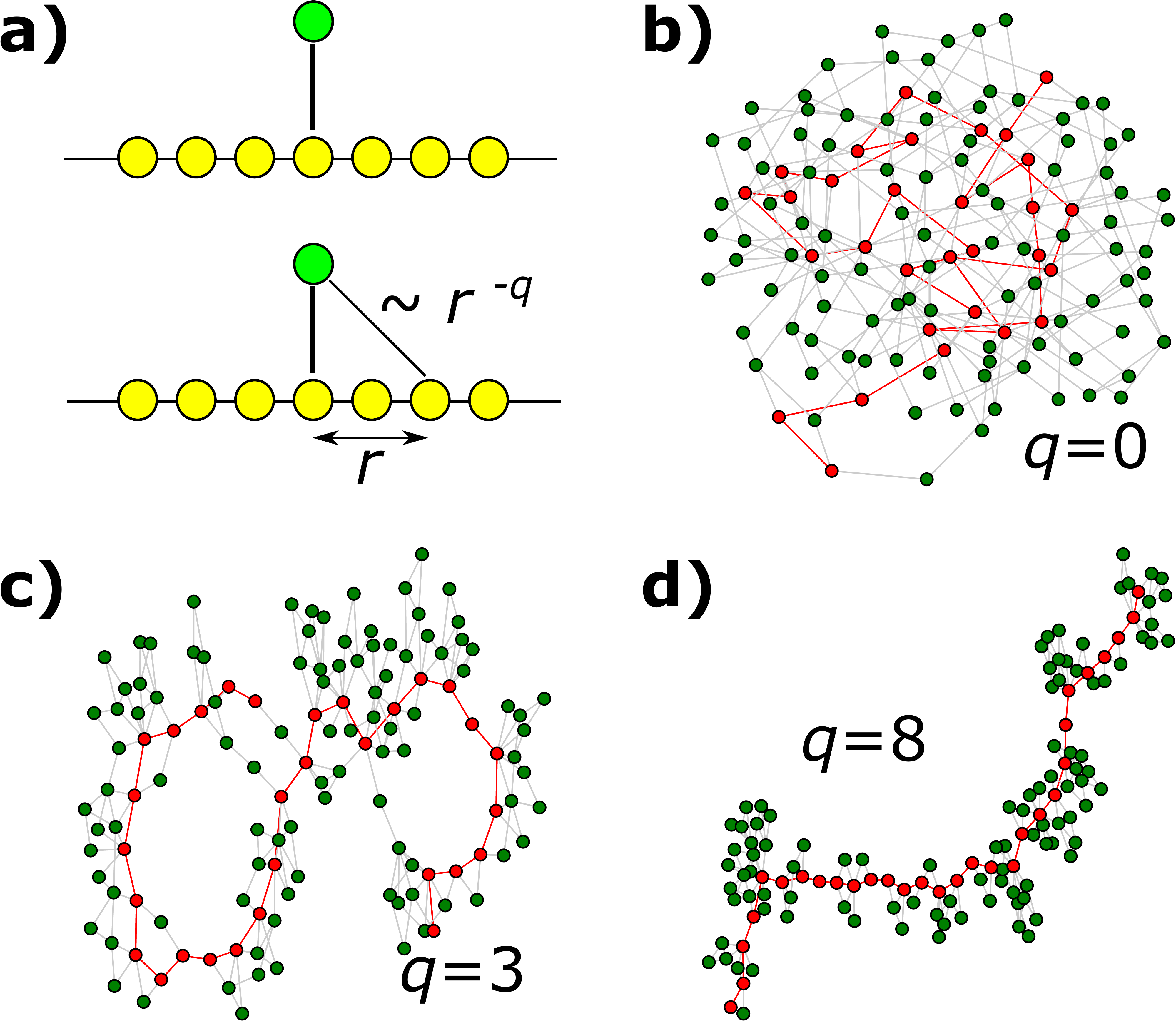}
\caption{The propinquity model. (a) The propinquity model can create networks with varying link density at different scales. The network grows via the successive addition of nodes (green) which link to a randomly selected existing node (thick line). The green node then selects a new node with probability $r^{-q}$, where $r$ is the distance from the previously selected neighbor. The network topology is controlled by varying the value of the parameter $q$. (b-d) Examples of small networks ($N=130$) created by varying the parameter $q$ in the propinquity model. The seed network includes 30 nodes in a line, which are shown in red, and 100 nodes are added, shown in green, according to the propinquity strategy with $m=2$ links. The structural differences are evident as we increase $q$ from $q=0$ (random recursive network) to larger values, such as $q=8$ (where new nodes remain locally connected and always form a triangle with two existing neighbors).}
\label{fig2}
\end{figure}

To model the varying strength of preference as a function of the network distance, we start with a small seed network of e.g. $m$ nodes connected to each other (indeed, any possible configuration of a connected network does not influence the results). The network grows by the introduction of a new node $i$ at each time step, when $i$ creates m links towards the existing network. The first link is created randomly by choosing a node $j$ (either preferentially or uniformly). The new node $i$ then creates $m-1$ additional links, where a node $h$ is now selected to be connected to node $i$ with probability $r_{jh}^{-q}$. The distance $r_{jh}$ denotes the shortest distance in the existing network between nodes $j$ and $h$, and $q$ is a parameter that controls how close the new connections will remain to the first choice. A schematic description of the algorithm is shown in Fig.~\ref{fig2}a. In Figs.~\ref{fig2}(b)-(d) we present some typical structures resulting from this algorithm for $m=2$, as we vary the value of $q$. The random character of the network at $q=0$ starts to break as we increase $q$, and fewer large-scale loops remain. For large values of $q$ the new nodes attach only to neighboring nodes, and the linear character of the network is preserved, with no long-range loops, see Fig.~\ref{fig2}d.

This model can describe a number of realistic situations. For example, new members that are invited into a social network will most likely connect to the close neighborhood of the member who invited them, and in spatially embedded networks, cost optimization makes shorter links preferable. Similarly, in co-purchase networks, if two items are frequently bought together, there is a larger probability that a buyer will prefer a new item in the same category \cite{36}, which will remain within the extended neighborhood of these items. In this way, we assume that new connections favor to remain close to already existing connections of the same node (hence, `propinquity'). Even beyond the realms of association as individual choice, biological networks result from the gradual accrual of small mutations that alter functional pathways one change at a time. Altering the viability of an organism one mutation at a time can similarly be considered as a propinquity-driven process with the potential to explain dynamics of conserved complexes \cite{hirsh2007identification} and offer novel, foundational frameworks for consideration of such network behaviors for applications including developmental biology \cite{FischerSmith2012} and drug discovery \cite{mestres2008data}.

The limiting cases $q=0$ and $q=\infty$ correspond to random selections over the entire network and strictly neighboring selections, respectively. As $q$ increases we expect that the model will result in an increasingly modular structure, since the links remain local and there are very few links that connect distant parts of the network. At the same time, the value of $q$ controls the local density scaling, with direct impact on network topology.

\section*{Results}

\subsection*{Results of the model}

We have studied two main variants of the model, which differ in the attachment mechanism of the first connection. In the first variant, a new node selects its first connection randomly, while in the second variant, the selection is preferential, i.e. proportional to the degree of an existing node. It is quite straightforward to calculate the degree distribution for the limiting cases of both variants (see Supporting Information). For random attachment, the distribution of the degree $k$ goes from exponential at $q=0$, $P(k)\sim(1+1/m)^{-k}$, to a power-law distribution $P(k)\sim k^{-\lambda}$ with an exponent $\lambda=2m+1$, i.e. $P(k)\sim k^{-(2m+1)}$, for large $q$. For preferential attachment, the degree distribution remains a power-law with an exponent changing from $\lambda=2m+1$ at $q=0$ to an exponent $\lambda=3$ at large values of $q$ (where the propinquity model becomes similar to a growing Barabasi-Albert model \cite{12}). Notice that for $m=1$ the exponent is $\lambda=3$, i.e. the propinquity model generalizes the BA network generation method. Critically, even though the two variants (random first selection with $q=8$ and preferential first selection with $q=0$) lead to the exact same degree distribution they are structurally different. In the first case, we select a random node and the second selection connects to a neighbor of the first node, which leads to an effective preferential attachment mechanism for the second choice, where the network evolves by forming new triangles leading to a large clustering coefficient. In the second case, the first node is selected preferentially and the second node is selected randomly, so that the number of triangles (and therefore the clustering coefficient) is practically zero. In this example, the global link density and the degree distribution are identical, so the clustering coefficient can be used to separate these two cases. However, the clustering coefficient only counts loops of 3 nodes and by varying $q$ we can find examples where loops of larger sizes are favored over triangles, while the clustering coefficient is still very close to 0. The networks in this case seem statistically similar under most of the standard network measures, masking their fundamental differences in local density.

In the current study, we calculate the dependence of $\rho_n$ and $t_n$ (Eqs.~(\ref{Eq1}) and (~\ref{Eq2})) on the sample size, $n$, by randomly sampling different parts of the network and averaging over the samples (see SI for details). We studied the possible scaling of $\langle t_n \rangle$ vs $n$ and found that, typically, we recover a power-law behavior. This power law form is described by the value of the exponent, $x$, in
\begin{equation}
\langle t_n \rangle \sim An^{-x} .
\label{Eq3}
\end{equation}

This scaling is more prominent for smaller values of $n$, when the subgraph size is significantly smaller than the network size, $N$. This is since our approach, due to the attractive interaction between successive links, is sensitive to local topologies where $n\ll N$. As we increase $n$, there is a crossover point after which $\langle t_n \rangle$ decays much faster with $n$, typically as $\langle t_n \rangle \sim n^{-1}$. This approximate pattern is true for most cases that we studied, but the exact behavior of $\langle t_n \rangle$ can vary depending on the structure.

Equation (\ref{Eq3}) describes how the density of links changes as we increase the scale of observation, through the value of the exponent $x$. If $x$ is close to 0, this means that the $n$-tangle density remains constant at any size, while for larger values of $x$ the density decays faster suggesting that larger areas of the network tend to become more tree-like. The variation of the exponent makes it also possible to monitor a possible transition of the structure in a given scale, from a tree to a denser graph, or vice versa. Notice that the magnitude of the density is controlled by the value of the prefactor, $A$, independently of the scaling with the size.

The calculation of the exponent $x$ is straightforward for simple structures, such as ER networks and lattices (see SI). In ER networks, there is no variation of the density with $n$, so that $x=0$. In lattices, as we discuss in the next section, the asymptotic value of the exponent is $x=1$. In general, the exponent $x$ can vary between 0 and 1, and therefore the lattice and the random network are representative of two extreme behaviors of how density can scale with size. Clearly, this means that we can characterize networks in this way as being closer to, or further from, particular structures, such as in the case of lattice or random networks\cite{27}. Notice that using the standard definition of local density $\rho_n$ in Eq. (\ref{Eq1}), we always retrieve the trivial behavior $\rho_n \sim n^{-1}$, which does not carry any useful information on local density.

\begin{figure}
\centering
\includegraphics[width=.95\linewidth]{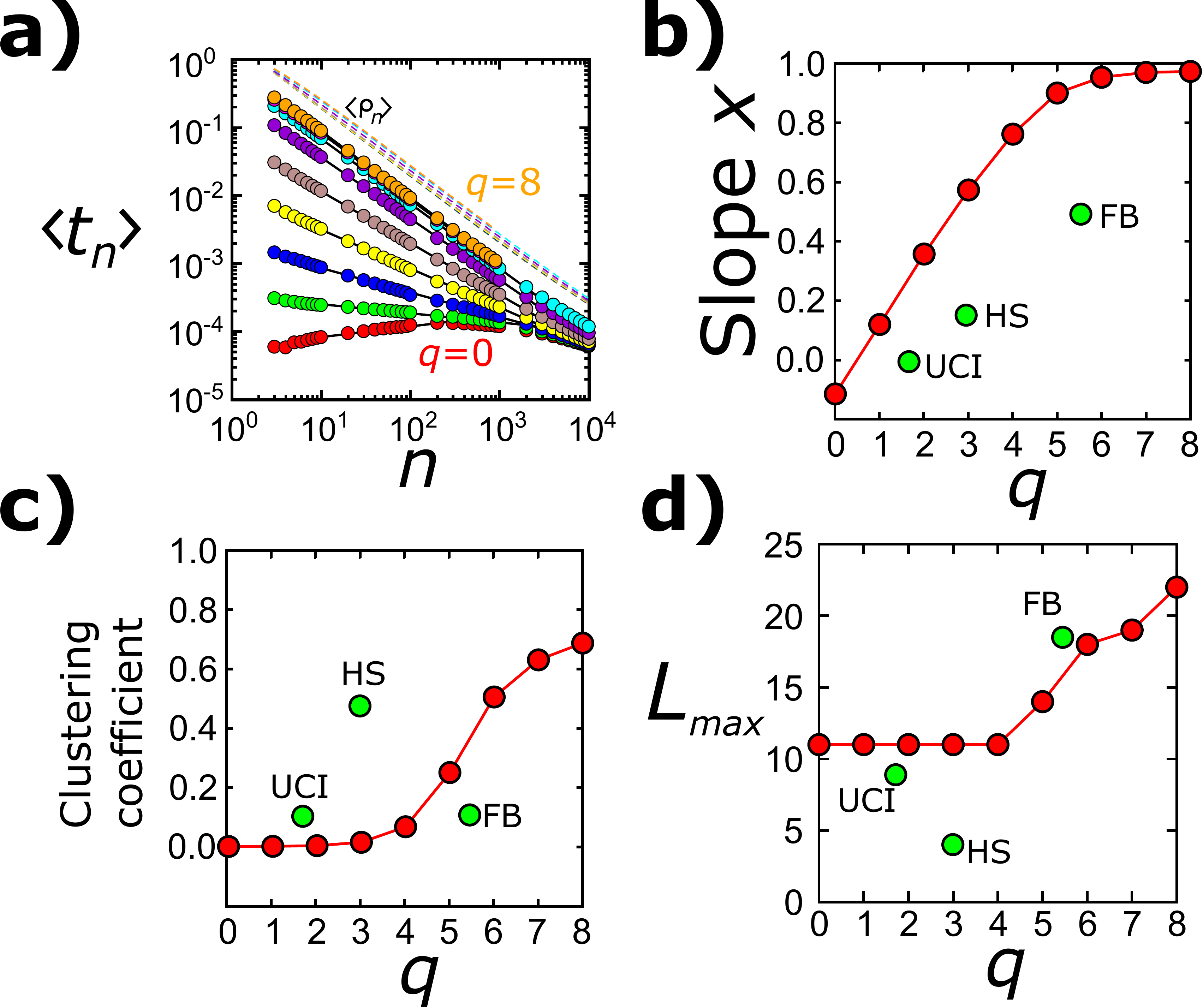}
\caption{Results for the propinquity model. Here, the first link of a new node attaches preferentially to the existing network. (A) Scaling of $n$-tangle density as a function of $n$. From bottom to top, the value of $q$ increases from 0 to 8 in steps of 1. Dashed lines correspond to regular density $\langle \rho_n \rangle$, where there is no observable effect of $q$ (the slope remains constant). (B) Calculation of the exponents $x$ for the lines in panel (b), as a function of $q$. The green circles indicate the corresponding values for the empirical networks analyzed in Fig.~\ref{fig1}. (C) Clustering coefficient as a function of $q$. (D) Even though the exponent $x$ increases with $q$, the network diameter (as well as the clustering) in the propinquity model remains constant up to $q=4$, and increases only for larger values of $q$.}
\label{fig3}
\end{figure}

We used the case of $m=2$ links per new node. As expected, when $q=0$ the connections are all random and we recover the result for random ER networks, where $t_n$ does not change significantly with $n$. As we increase the value of $q$ the density starts to change systematically with $n$, following a power law behavior (Fig.~\ref{fig3}a). This is reflected in the value of the exponent $x$ which starts at $x=0$ when $q=0$ and increases monotonically until it reaches values close to $x\sim 1$ (Fig.~\ref{fig3}b).

Interestingly, while the local density changes drastically with $q$, and we can therefore deduce that large structural changes take place, we would not be able to observe these changes by using standard network measures, such as clustering and distances. In Fig.~\ref{fig3}c, the clustering coefficient remains almost 0 for values between $q=0$ to $q\sim 4$, but the local density behavior is drastically different, as can be seen in the results of Fig.~\ref{fig3}a and the slope calculations in Fig.~\ref{fig3}b. Similarly, the network diameter remains unchanged in the range of $q$ from 0 to 4 (Fig.~\ref{fig3}d). In the same $q$ range, the slope of the density increases from 0 to 0.6. These results show that even though the relative distances remain constant, the links reorganize themselves in a systematic way with larger local densities at small subgraphs. The local link density exponent can therefore be used to characterize changes in network structures that cannot be predicted by the study of the clustering coefficient or shortest paths. When q assumes large values, both the clustering coefficient and the network diameter increase significantly as a result of highly localized connections and the removal of practically all network shortcuts. However, in this range there is very little variation in the local density, $t_n$, see Fig.~\ref{fig3}b.

\subsection*{Real networks}

In real systems, when a node creates a new link there are obviously many possible mechanisms in action, e.g. homophily and collective action \cite{37}, consensus dynamics \cite{lu2009naming}, etc.  The propinquity model, however, allows us to isolate the influence of neighbor's proximity to network density. It then provides a simple model by which to predict the variation of link density at different scales, even though the use of the typical link density definition would falsely indicate that the extent of the propinquity concept (through the parameter $q$) should have no influence on the results.

\begin{figure}
\centering
\includegraphics[width=.9\linewidth]{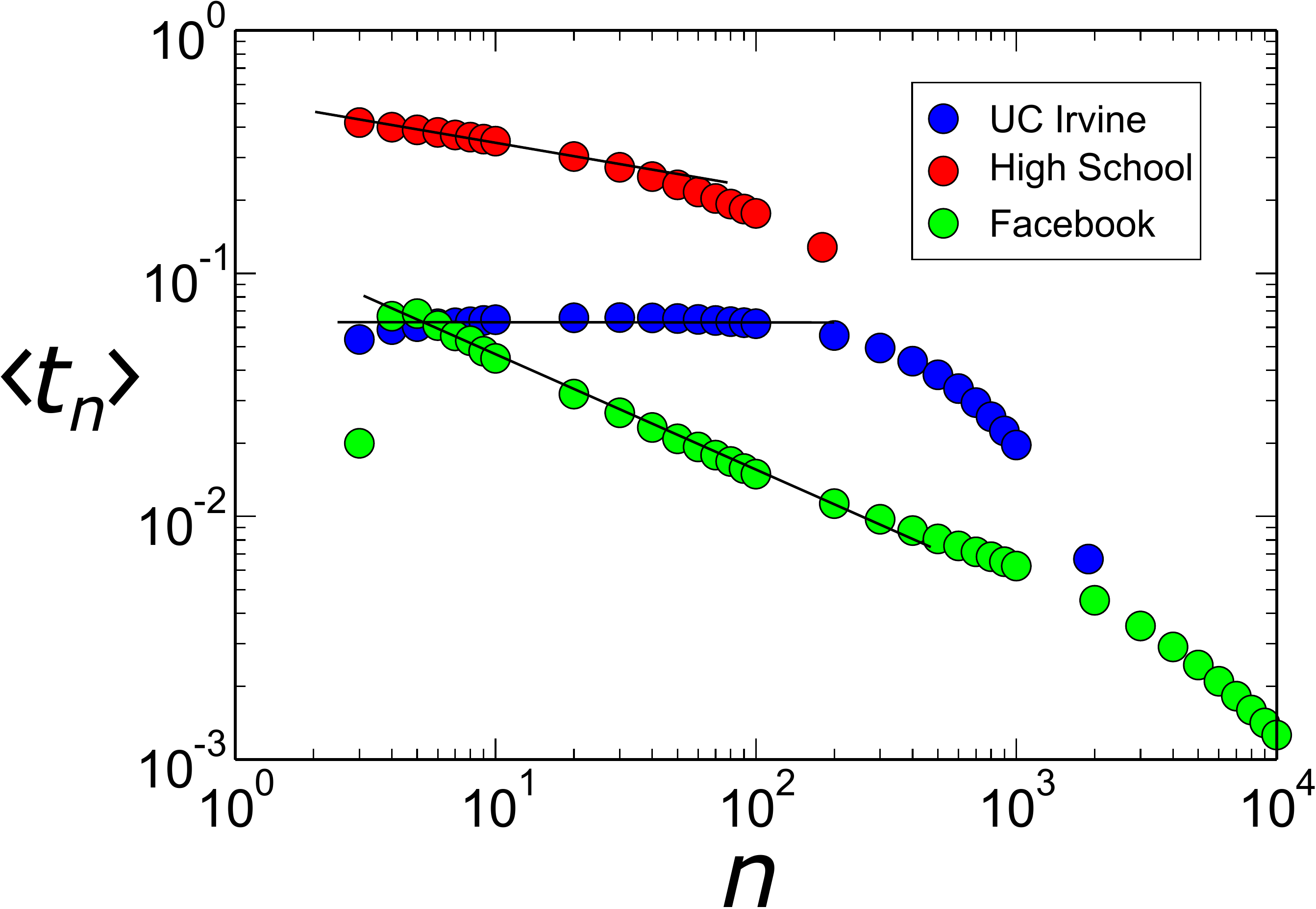}
\caption{Density scaling of the real networks in Fig.~\ref{fig1}. The exponent, $x$, that characterizes the density scaling in the empirical networks of Fig.~\ref{fig1} is consistent with the propinquity model exponents $q$. These exponents ($x$=0, 0.16, and 0.5) are shown in Fig.~\ref{fig3}c and follow the same trend, increasing with $q$, as the propinquity model in that plot.}
\label{fig4}
\end{figure}

In Fig.~\ref{fig4} we calculate the $n$-tangle density scaling for the three empirical networks analyzed in Fig.~\ref{fig1}. Each network leads to a different slope, $x$. Using the optimal value for $q$ from Fig.~\ref{fig1} and the exponent $x$ from Fig.~\ref{fig4}, we can compare the propinquity metrics for these networks. Of course, as mentioned above, the empirical data cannot be assumed to be fully described by one mechanism alone. However, it is clear from Fig.~\ref{fig3}b that there is a consistent trend in both the model and in empirical data that larger local density variations appear at larger $q$ values. This observation is important because it provides a link between the analysis of a static network snapshot and the network generation mechanism, which is difficult to observe directly. In practice, we have shown that measurements of the scaling of local link density provide a systematic way to understand network growth mechanisms which are based on the distance between two nodes, added one after another as friends.

As a comparison, network properties such as the clustering coefficient or the network diameter (shown in Figs.~\ref{fig3}c-d) do not suggest any clear trends with $q$. However, this may also be attributed to the small size of these networks, such as the High School network, which only contains 180 students and is an unusual, dense network.

\section*{Discussion}
Our work demonstrates the importance incorporating mechanisms of attachment that allow the tailoring of local network densities to achieve realistic network structures in generative growing models. We have studied the simple case where the second link depends on the in-network distance and we have shown that this leads to very different topologies. This finding was confirmed by studying the distance between the first two neighbors of new nodes in empirical networks.

We establish a family of network generation models where the subsequent connections depend on the distance between two nodes. To detect the influence of this mechanism on topology we study the scaling of local density. If we use the standard definition of local density then the scaling is dominated by a trivial structure. However, we show that a re-definition of local density, Eq. (\ref{fig2}), provides a direct way of studying this scaling and the local density can probe the structure at different scales.

From a theoretical point of view, the power-law behavior in Eq. (\ref{fig3}) can also be seen as the definition of a new fractal dimension for complex networks, albeit within the range from $x=0$ to $x=1$. The traditional definition of a fractal object detects how the mass scales with distance. In complex networks, this definition becomes problematic because of the natural restriction of distances in usually just one decade. For example, the maximum distance in the three empirical networks used in Fig. 1 ranges from 4 to 19, which does not allow a reliable evaluation of network dimensions (see also related discussion in the SI). There are many methods in the literature that have introduced possible modifications on how fractal features can be measured in networks \cite{song2007calculate,li2011dimension} , but even then there are many non-fractal networks (e.g. Erdos-Renyi networks) whose structural differences cannot be captured by fractal dimension. As an alternative to these methods, the present link density method can provide a natural interpretation of the self-similar properties of a network. In this definition, the important quantity is the `mass' of the links instead of the number of nodes (see also \cite{12}), while the `length' corresponds to the number of nodes, instead of a distance metric. Self-similarity in this study shows how the fraction of the excessive links scales with the number of nodes. A small exponent means that any part of the network will have similar link density, independently of the sampled size, but a large exponent shows that larger samples of the network become sparser. The rate at which the density decreases is then determined by this fractal exponent $x$.

In conclusion, the propinquity model provides a new class of generative models, rooted in features of real networks and is leading to understanding how individuals become integrated into communities at different scales. It enables us to test meaningful hypotheses about which scales of social interactions are important in an evolving network as a metric for isolated analysis and comparison between systems. Most importantly, it allows us to make behaviorally-driven predictions about the emergent structure of networks based on single snapshot observations.

\acknowledgments We acknowledge support by the National Science Foundation (NSF) under Grants CNS-1646856 (to L.K.G.) and CNS-1646890 (to N.H.F.). S.H. acknowledges financial support from US/Israel Binational Science Foundation (NSF/BSF) No. 2015781; Israel Science Foundation (ISF); Office of Naval Research (ONR) Grant N62909-14-1-N019; the Israel Ministry of Science and Technology with the Italian Ministry of Foreign Affairs; the Israel Ministry of Foreign Affairs with the Japan Science Foundation; and the Army Research Office. H.E.S. and S.H. acknowledge financial support from Defense Threat Reduction Agency (DTRA) Grant HDTRA-1-14-1-0017.

\bibliography{Prop_paper}

\newpage

\textbf{\Large Supplementary Information for: Propinquity drives the emergence of network structure and density}

\section*{Empirical networks}
The study of second links in real networks requires knowledge of their
complete time evolution. There are very few available datasets in the
literature which fulfill this requirement. The three networks that we
used in this work are as follows:

- \emph{UC Irvine}. We downloaded this dataset from
\url{http://toreopsahl.com/datasets/\#online_social_network} . This
network has been studied in detail in Ref. (17). The dataset describes
interactions, in the form of online messages in a Facebook-like setting,
between students at the University of California, Irvive. These messages
are timestamped and so we were able to reconstruct the order by which
this network was built. The network includes 1893 nodes and 13835 links.

- \emph{High School}. This dataset was downloaded from
\url{http://www.sociopatterns.org/datasets/high-school-dynamic-contact-networks}
and contains the temporal evolution of contacts between students in a
High School in Marseilles, France, over a 5 day period in November 2012
(18). The network includes 180 nodes and 8145 links.

- \emph{Facebook wall}. This dataset was downloaded from
\url{http://socialnetworks.mpi-sws.org/data-wosn2009.html} . The data
contain the evolution of the link structure in the Facebook New Orleans
networks (19). The links represent communication through the wall
feature of Facebook. The network includes 43953 nodes and 182384 links.

\section*{The problem with density measurements}

The basic quantity that we study in this work is the link density. We
are mainly interested in determining the behavior of local link density,
and how this varies at different scales within the network. The scale is
determined by the number of nodes, $n$, in a connected subgraph of
the network (which itself frequently represents a subsampled graph of a
larger network). Formally, the link density $\rho$ in a graph with
\emph{N} nodes and \emph{L} links is defined as the fraction of the
number of links over the maximum possible number of links, i.e.
$\rho$=\emph{L}/{[}\emph{N}(\emph{N}-1)/2{]}. To measure the local
link density we consider an induced connected subgraph of \emph{n}
nodes, where we take into account all the \emph{e\textsubscript{n}}
existing links between all pairs of nodes in the subgraph {[}this
subgraph closely resembles the outcome of typical BFS sampling
methods{]}. We then define the local link density as
\begin{equation}
\rho_{n} = \frac{e_{n}}{\frac{n(n - 1)}{2}} .
\end{equation}

How does the density scale with the number of nodes? The answer should
be particularly simple in, e.g., ER networks, where we know that the
structure is homogeneous and there are diminishing fluctuations of the
density in any part of the network. The subgraph size should not
influence the density measurement, and any subgraph should yield similar
link density. To measure such a quantity there are traditionally two
approaches which are considered fully equivalent to each other. First,
we can create ER networks of different size \emph{N} and measure how the
density scales with \emph{N.} As shown in Fig. S1a, this is a trivial
computation where $\rho$ is constant, independently of the size \emph{N.} The
second approach is to consider a large network of \emph{N} nodes and
randomly sample smaller connected subgraphs of \emph{n} nodes, and
proceed by varying \emph{n}. In this case, however, the density is no
longer constant, but decays roughly as \emph{n}\textsuperscript{-1}
(Fig. S1b), before reaching its asymptotic expected value as in Fig.
S1a. This means that if we did not know anything about this network and
we were only sampling small parts of it, we would not be able to deduce
that these subgraphs were parts of an ER network.

\begin{figure}
\centering
\includegraphics[width=.95\linewidth]{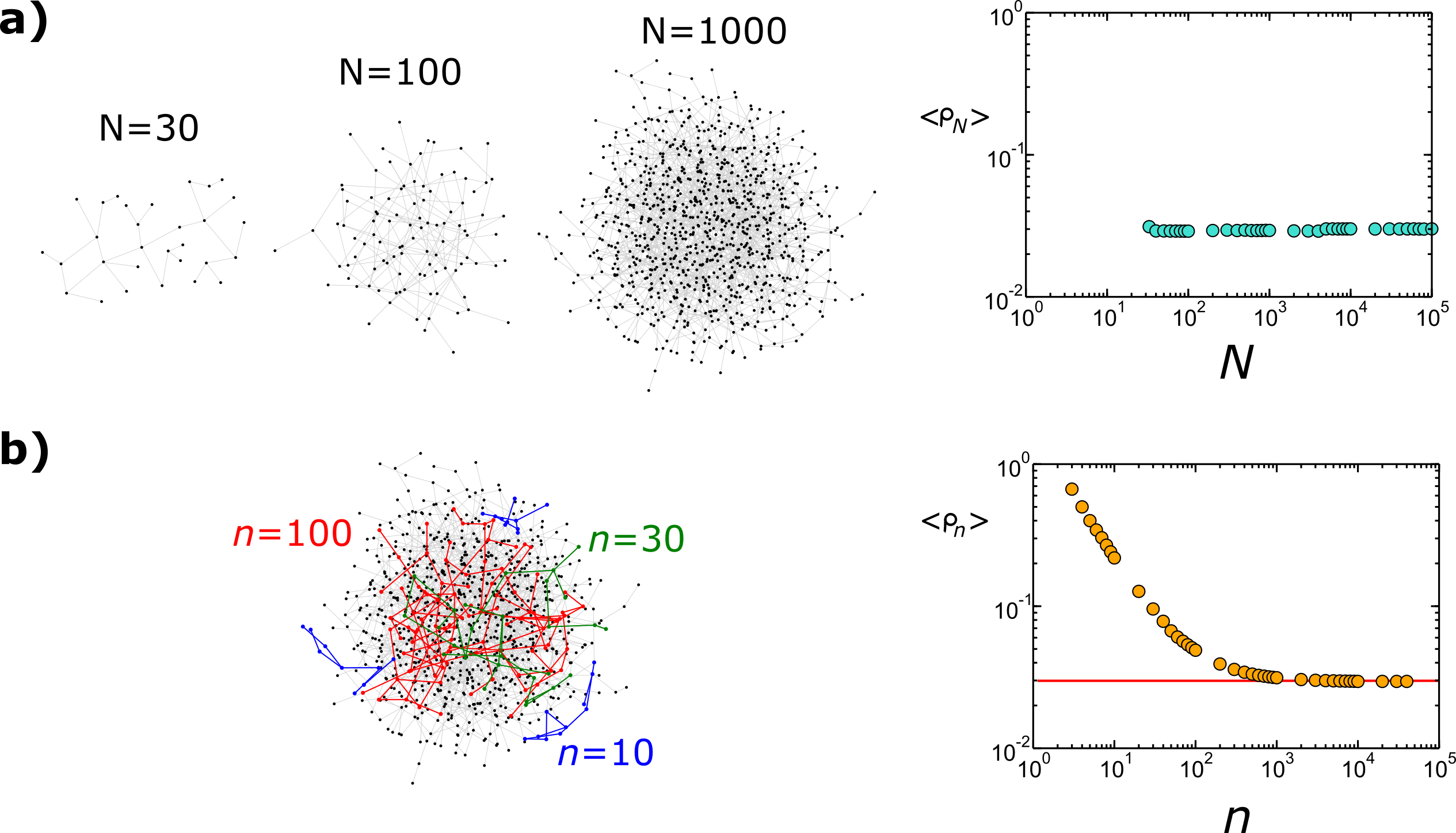}
\caption{
\textbf{The problem with link density across networks and
within a network. (a)} When we construct ER networks of different size,
$N$, with the same link density, $\rho$, ~(here $\rho$=3 10\textsuperscript{-2}),
then the observed link density trivially remains constant independently
of the network size $N$, as expected. The same is trivially true for
scale-free networks, or any other model structures where the global
network density is fixed by the model parameters. Notice that it is not
possible to construct a connected network with size smaller than $1/\rho$ (so here, N\textgreater{}33). \textbf{(b)} In contrast, the link density
in connected subgraphs of size $n$, within a network of size $N$, decays
inversely linearly with $n$ when the subgraphs are sampled from a larger
network, before reaching its asymptotic value. In the plot, we sample
subgraphs from an ER network of N=10\textsuperscript{5} nodes, where the
global link density of the network is also equal to $\rho$=3
10\textsuperscript{-2}.}
\end{figure}

This behavior is due to the fact that choosing randomly \emph{n} nodes
from an \emph{N}-nodes graph (\emph{n}\textless{}\textless{}\emph{N})
yields subgraphs that are below the percolation threshold, and therefore
the number of links in such a subgraph should be less than \emph{n},
i.e. the subgraph cannot be connected. Thus, the sampling process
applied here is biased, i.e., it always provides connected subgraphs,
which means that significant correlations are introduced. If we were
sampling a real network, we would not be able to know in advance the
minimum sampling size required for the given density, since the global
density remains unknown unless we can have access to the entire network.
The scaling of the subgraph density with the size would then mislead us
to conclude that we were sampling a non-random network, where density is
not constant at different scales.

The simple solution that we suggest here is to subtract these \emph{n}-1
links in the tree structure that introduce artificial correlations. The
snowball sampling method and the definition of the connected subgraph
require all the subgraphs to have at least \emph{n}-1 links for
connectivity. As mentioned above, in real networks the density has been
shown to scale inverse linearly with the network size, i.e.
$\rho_n \sim n^{-1}+O(n^{-2})$.
This means that
$e_n\sim n + O(1+n^{-1})$
and the linear term dominates the behavior of $e_n$.
If we simply subtract these links from \emph{e\textsubscript{n}}, the
density behavior will now depend on the higher-order terms, which may
scale in a different way than $\rho_n$. We therefore
apply here the recently defined metric for the local \emph{n}-tangle
(Topological Analysis of Network subGraph Link/Edge) density,
\emph{t\textsubscript{n}}, as

\begin{equation}
t_{n} = \frac{e_{n} - (n - 1)}{\frac{n(n - 1)}{2} - (n - 1)}
\end{equation}

The key feature in this definition is the removal of the \emph{n}-1
links that are necessarily present in an induced subgraph to ensure
connectivity. We also remove the same number of links in the
denominator, so that \emph{t\textsubscript{n}} remains properly
normalized and ranges from \emph{t\textsubscript{n}}=0 in the case of a
loopless tree subgraph to \emph{t\textsubscript{n}}=1 for a fully
connected subgraph.

In the current study, we calculate the dependence of $\rho_n$
 and $t_n$ on the sample
size, \emph{n}, by randomly sampling different parts of the network and
averaging over the samples. In practice, we start by fixing the number
of nodes \emph{n}. We select a random node and add it to the subgraph.
We then create a list that includes all the links of this node and
randomly select one of these links. The node at the other end of the
link is added to the subgraph and its links are added to the candidate
list. We repeat this process until the subgraph includes \emph{n} nodes.
Finally, we add all the links between the subgraph nodes that appear in
the original network, creating thus the induced subgraph. We calculate
the number of links \emph{e\textsubscript{n}} in this subgraph and
convert it to $\rho_n$ and \emph{t\textsubscript{n}}
according to equations [1] and [2]. We repeat this procedure and build
the corresponding distributions, which finally yield the average values
$\langle \rho_n \rangle$ and $\langle t_n \rangle$. We can then change
the value of \emph{n} and generate subgraphs of different size.

We studied the possible scaling of $\langle t_n \rangle$
 vs $n$ and found
that, typically, we recover a power-law behavior. This power law form is
described by the value of the exponent, \emph{x}, in
\begin{equation}
\left\langle t_{n} \right\rangle\sim An^{- x} .
\end{equation}

This scaling is more prominent for smaller values of \emph{n}, when the
subgraph size is significantly smaller than the network size, \emph{N}.
As we increase \emph{n}, there is a crossover point after which
$\langle t_{n} \rangle$ decays much faster
with \emph{n}, typically as
$\langle t_{n} \rangle$\textasciitilde{}\emph{n}\textsuperscript{-1}.
This approximate pattern is true for most cases that we studied, but the
exact behavior of $\langle t_{n} \rangle$ can
vary depending on the structure.

Equation [3] describes how the density of links changes as we increase
the scale of observation, through the value of the exponent \emph{x}. If
\emph{x} is close to 0, this means that the \emph{n}-tangle density
remains constant at any size, while for larger values of \emph{x} the
density decays faster suggesting that larger areas of the network tend
to become more tree-like. The variation of the exponent makes it also
possible to monitor a possible transition of the structure in a given
scale, from a tree to a denser graph, or vice versa. Notice that the
magnitude of the density is controlled by the value of the prefactor,
\emph{A}, independently of the scaling with the size.

The calculation of the exponent \emph{x} is straightforward for simple
structures, such as ER networks and lattices. In ER networks, there is
no variation of the density with \emph{n}, so that \emph{x}=0. In
lattices, as we discuss in the next section, the asymptotic value of the
exponent is \emph{x}=1. In general, the exponent \emph{x} can vary
between 0 and 1, and therefore the lattice and the random network are
representative of two extreme behaviors of how density can scale with
size. Clearly, this means that we can characterize networks in this way
as being closer to, or further from, particular structures, such as in
the case of lattice or random networks.

\section*{Calculation of subgraph density in Erdos-Renyi networks}

We consider \emph{n}-nodes in an induced connected subgraph of a larger
Erdos-Renyi network. We denote the probability for any link to exist in
this network as \emph{p.} The total number of possible links in the
subgraph is \emph{n}(\emph{n}-1)/2. Since the subnetwork is connected,
there are already at least \emph{n}-1 links in the subgraph. Each of the
remaining \emph{n}(\emph{n}-1)/2-(\emph{n}-1) possible links appears
with probability \emph{p}. As a result, the total number of links in the
subgraph, \emph{e}\textsubscript{n}, is the sum of these two quantities,
i.e.

\begin{equation}
e_{n} = ( n - 1) + p \left( \frac{n(n - 1)}{2} - (n - 1) \right) .
\end{equation}

Using the definitions in Eqs. (1) and (2) of the main text, it is easy
to show that Eq. [4] yields the following results:
\begin{equation}
\langle t_{n} \rangle=\emph{p} \;\;,\;\;
\langle \rho_n \rangle= p + \frac{2(1 - p)}{n} .
\end{equation}
The key idea in this calculation is that the definition of a connected
subgraph already imposes the existence of \emph{n}-1 links.

\section*{Density measurements in model and real networks.}

We first study the behavior of density in simple model network
structures, where we already have an intuition for how links are
organized. In Fig. S2a we show the dependence of both the regular and
the \emph{n}-tangle density, as a function of the subgraph size
\emph{n}, for ER networks of varying global densities. It is clear that
in all ER networks there is almost no change of the
\emph{t\textsubscript{n}} density for any value of \emph{n}, i.e. the
exponent \emph{x}=0, independent of the network density. The value of
\emph{t\textsubscript{n}} is trivially equal to the average density of
the global network. As noticed above, the regular definition of density
leads to a power-law decay with exponent \emph{x}=1, instead, and
constant density is recovered only asymptotically. The exact calculation
of these densities in ER networks (Eq. [5]) shows that these two quantities behave differently, but this difference is not a universal
feature. For example, in regular square lattices (Fig. S2b) the behavior
of both measures is the same, and they both scale as
n\textsuperscript{-1}. It is quite simple to explain why this happens: a
subgraph of \emph{n} nodes in a lattice includes 2\emph{n} links out of
a maximum possible of \emph{n}\textsuperscript{2}, yielding a dependence
of 1/\emph{n}. Except for very small values of \emph{n,} when we
subtract the \emph{n}-1 links we only modify the prefactor of the ratio,
but not the scaling. This also corresponds to our intuition that larger
lattices are more diluted since the number of links increases linearly
with the number of nodes, but the number of possible links is
proportional to the square of this number, so that the density vanishes
asymptotically with increasing size.

In scale-free networks (Fig. S2c) we find that the n-tangle density
$\langle t_{n} \rangle$ remains roughly
constant for small values of \emph{n}, and asymptotically it decays
inversely linearly with \emph{n}. The constant value is significantly
higher than the global density, which here was 3 10\textsuperscript{-6},
indicating that there is a larger concentration of links in smaller
regions, at least when the degree exponent $\gamma$~is smaller than 3.0. This
is a result of the inhomogeneous character of the scale-free structure,
as can be understood by the presence of hubs with higher probability and the increased number of links around
them. In other words, a sampling process that discovers new nodes by
following links, tends to over-estimate the presence of hubs because
they are selected more often. The regular density
$\langle \rho_n \rangle$, however, scales as a
power-law for the entire range of \emph{n}. A comparison of the
scale-free results with the ER results demonstrates that
$\langle \rho_n \rangle$ cannot separate the
two cases, since they both decay in a similar way. The \emph{n}-tangle
density, $\langle t_{n} \rangle$, on the
other hand, does not change in ER networks, but for scale-free networks
it is larger in small scales compared to its value at larger scales.

For larger values of the degree exponent $\gamma$\textgreater{}3, the n-tangle
density is significantly smaller than the global network density,
because the hubs are much weaker and the structure is much closer to a
tree topology. In this case, there are very few excessive links and Eq.
(2) indicates that $\langle t_{n} \rangle$
can only have small values, which vanish for a tree structure.

We finally studied a model of explicit modularity. We created networks
of largely isolated modules of \emph{M} nodes each, where a node in the
module had 0.99 probability of creating connections within the module
and 0.01 between modules. The density
$\langle t_{n} \rangle$ describes this
structure very accurately, remaining constant until \emph{n} is equal to
the number of nodes, \emph{M}, in the module. After that, the behavior
changed abruptly towards a power-law decay, until \emph{n}=\emph{N}
where the \emph{n}-tangle density becomes equal to the global density
used here, $\rho$=4 10\textsuperscript{-4}. Therefore, a change in the
behavior of density at different scales can also be used as a detection
method for modularity and for estimating typical module sizes. The link
density $\langle \rho_n \rangle$ exhibits a
transition at the same point, but it cannot capture the constant density
within a module, similar to the case of ER networks.

\begin{figure}
    \centering
    \includegraphics[width=.95\linewidth]{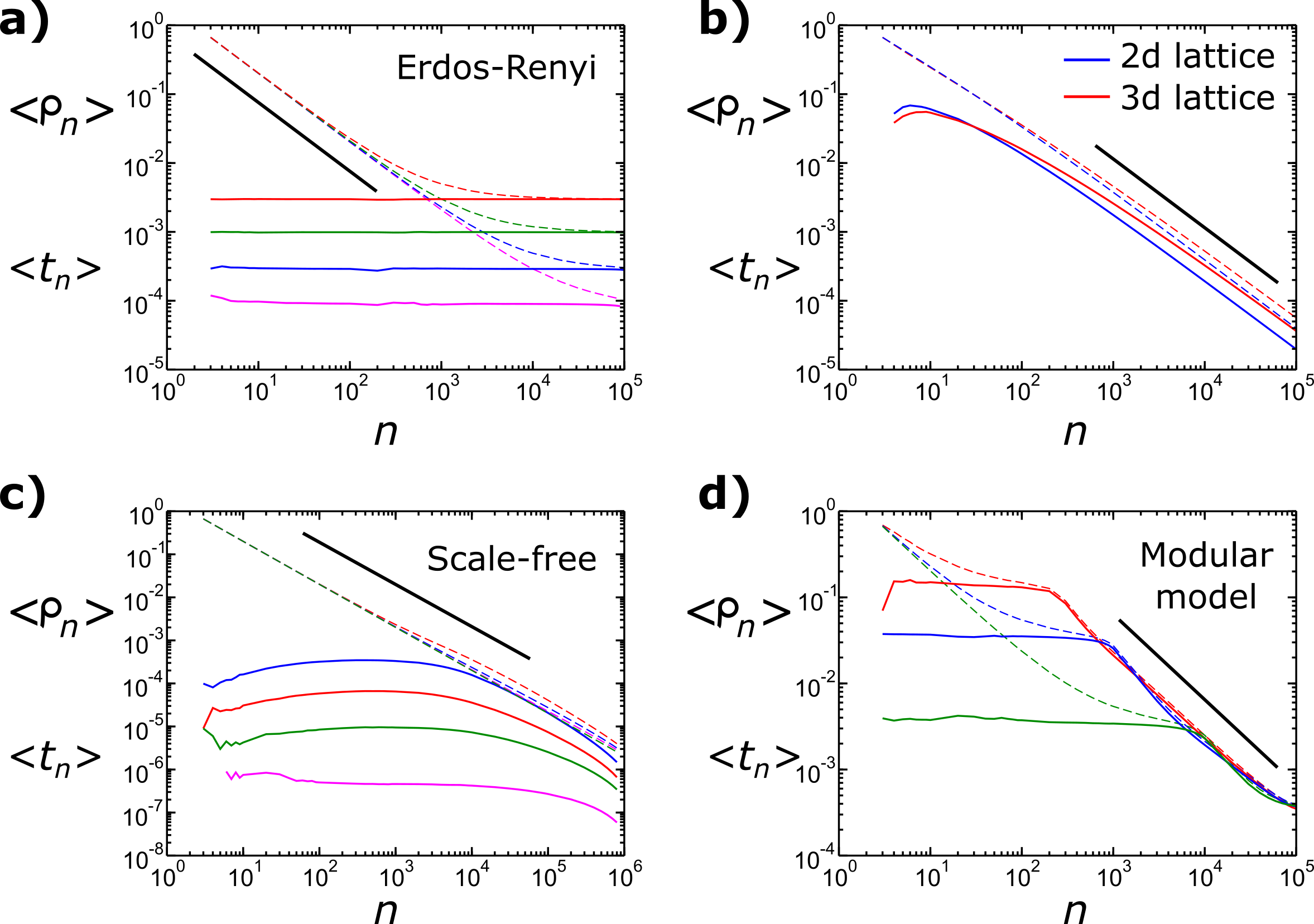}
    \caption{\textbf{Variation of the average n-tangle density $\langle t_n \rangle $ (continuous lines) and the
average link density $\langle \rho_n \rangle$ (dashed
lines), as a function of $n$, for typical network structures.} The black
lines are used as a guide to the eye for a slope equal to -1.
\textbf{(a)} Results for ER networks of size $N=10^5$ and
link densities (top to bottom):3x10\textsuperscript{-2},
10\textsuperscript{-3}, 3x10\textsuperscript{-3}, and
10\textsuperscript{-4}. \textbf{(b)} Results for two- and
three-dimensional lattices. \textbf{(c)} Results for scale-free
networks, created with the configuration model, with a degree exponent
(top to bottom): $\gamma$=2.5, 2.75, 3.0, and 3.5. The network sizes are
$N$=10\textsuperscript{6} and the global density is $\rho$ \textasciitilde{}3
10\textsuperscript{-6}. \textbf{(d)} Results for the modular model, for
networks of $N$=10\textsuperscript{5} nodes and modules with (top to
bottom): $M=$250, 1000, and 10000 nodes. Notice that the n-tangle density
remains constant until we reach the size of the module. The behavior
then changes and  $\langle t_n \rangle$ decays
almost inversely linearly. The crossover value in each line can be used
to estimate the typical size of the modules in the structure.
}
    \label{figS2}
\end{figure}

We also calculated the local density dependence on \emph{n} for a number
of real networks. These include: i) the Internet at the AS level (Caida
project) in four different years, ii) the Amazon co-purchase network at
three different dates, iii) the Gnutella sharing network, and iv) the
Facebook friendship network of US Universities in 2005.

These real networks present a range of different behaviors. For example,
the scaling of the \emph{n}-tangle density in the AS Internet (Fig. S3a)
behaves similarly to that of a modular structure, with a roughly
constant density up to n\textasciitilde{}1000 and an inversely linear
decay at larger sizes. We studied four different structures, separated
by one year between 2004 and 2007, and there was very little variation
in the density scaling, even though the structure itself has changed
over this time period. The
$\langle \rho_n \rangle$ density presents a
small transition range, which indicates that the absolute value of the
density changes abruptly, but retains the same scaling behavior for the
whole range of \emph{n}.

\begin{figure}
    \centering
    \includegraphics[width=.95\linewidth]{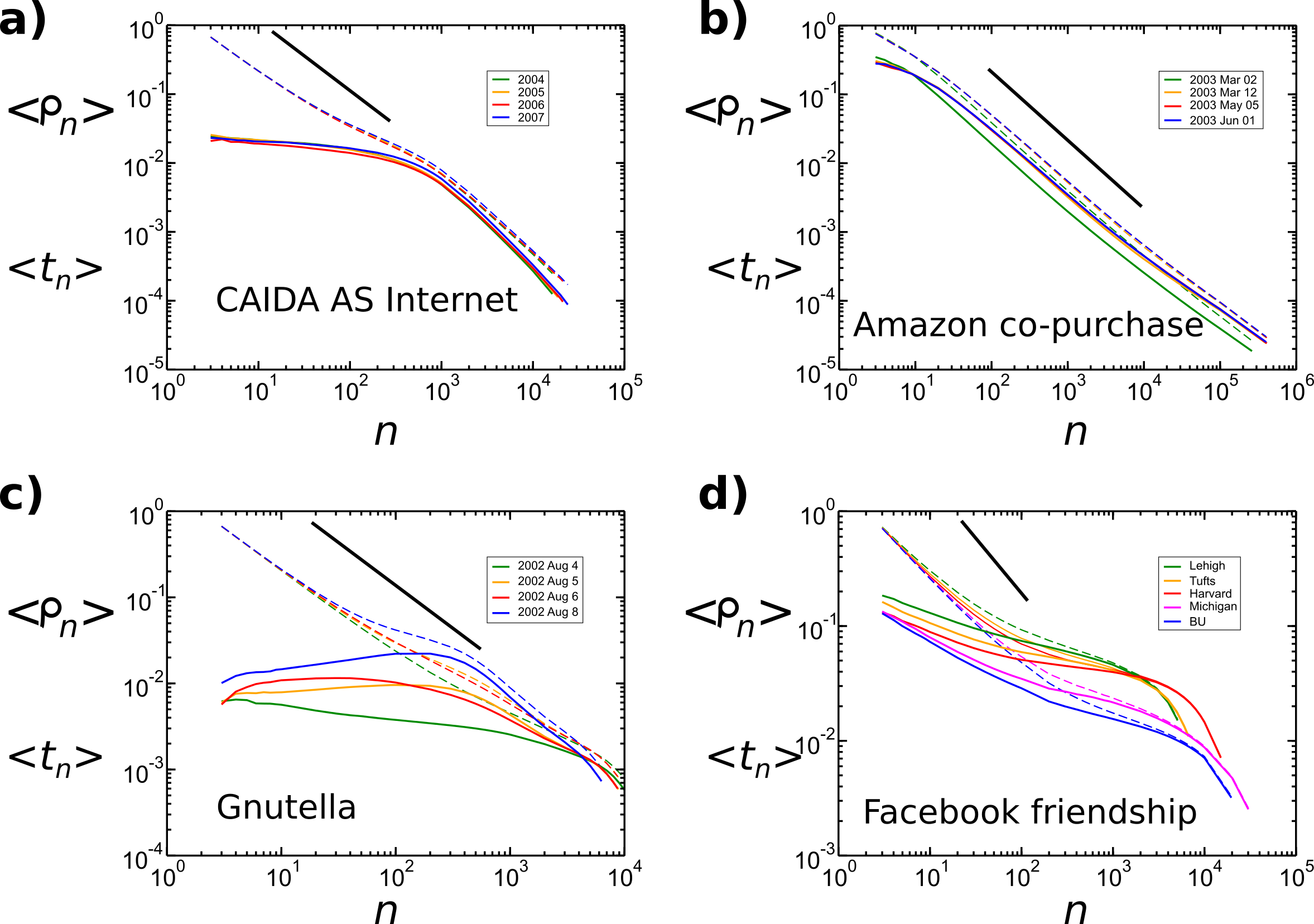}
    \caption{\textbf{Density scaling in real networks. (a)} AS Internet
connectivity at four different dates, from 2004 to 2007, \textbf{(b)}
Amazon co-purchase network at four different dates, \textbf{(c)}
Gnutella sharing network at four different dates, and \textbf{(d)}
Facebook friendship networks in 5 US Universities in 2005. Solid lines
represent $\langle t_n \rangle $, and dashed lines
represent $\langle \rho_n \rangle $.}
    \label{figS3}
\end{figure}

Surprisingly, all Amazon co-purchase networks (Fig. S3b) have density
features similar to that of spatial networks, such as the square
lattices of Fig. S2b. These networks are still scale-free with a broad
degree distribution, so spatial embedding in low dimensions is not
evident. Even though certain classes of scale-free networks have been
shown to be spatially embeddable under certain circumstances, the
majority of complex networks are difficult to embed and a high degree of
organization is required for a network to have features similar with a
low-dimension structure. It is puzzling, then, that in a network of this
size ($N\sim 400,000$) with a broad degree distribution, the
density would scale similarly with a lattice structure. This scaling
indicates a well-organized configuration of links, as we vary the size
of a subgraph. This regularity can be explained by the nature of the
connections. In the co-purchase network, two products, e.g. two books,
are connected when they are frequently bought together. This leads to a
significantly modular network, where books are highly connected within
their own category, e.g. fiction books, technical books, etc, and much
less across categories. This reflects the purchase habits of consumers,
who tend to be interested in items of just a few categories rather than
buying items with a uniform probability from among all categories. Our
results indicate that there is a large degree of order at all scales in
the structure, and links tend to remain local (like in a lattice), with
very few long-range shortcuts. This is in analogy with spatial link
arrangements, where larger subgraphs become significantly more diluted
and most links remain local.

The Gnutella p2p networks (Fig. S3c) exhibit a behavior reminiscent of
the random scale-free networks in Fig. S2c. At small values of \emph{n,}
we recover either a constant value of
$\langle t_{n} \rangle$ or a small decay,
i.e. \emph{x}\textasciitilde{}0.2. Asymptotically, this decay becomes
faster and the local density reaches much lower values. This variation
of the decay can be attributed to the strongly inhomogeneous character
of the structure, similarly with the case of the scale-free networks.
The hubs lead to an increased local density, but the average density of
the network as a whole is significantly lower.

Finally, in Facebook friendship networks (Fig. S3d), the networks that
we studied show a decay with the subgraph size \emph{n}, with moderate
exponents in the range \emph{x}=0.3-0.5. When we randomly rewired the
connections among nodes in these networks, keeping the degree of all
nodes intact, we found that the density remains constant with \emph{n}
and all exponents are very close to \emph{x}=0, as we would expect from
a random un-organized network.

\section*{Calculation of the degree distributions in the propinquity
model}

We studied two variations of the propinquity model. In the first case, a
node attaches randomly to an existing node and then uses the propinquity
principle to find its second connection. In the second case, the new
node selects its first connection preferentially, i.e. with a
probability that depends linearly on the existing degree of each node.
Using the standard technique of rate equations we can easily calculate
the degree distribution for the extreme cases of $q=0$ and
$q\rightarrow \infty$. In the following, we assume a growing network which starts at time
$t=0$ without any nodes. Let $N(k,t)$ be the number of degree $k$ nodes at
time $t$ and $N(t)$ the total number of nodes in the network. At every time
step, a new node is added to the network, so that $N(t)=t$. Every new node
connects to $m$ existing nodes, where the first node is selected either
randomly or preferentially and the remaining $m-1$ nodes are selected
depending on their distance from the first node, according to the
propinquity model. The probability that a node has degree $k$ at time $t$ is
then: $p_k(t)=N(k,t)/N(t)$.

\subsection*{a) Random selection of the initial node and \emph{q}=0}

This model corresponds to a new node which connects to $m$ random nodes in the
network. The probability to select a node is independent of the degree
or the network distance and it is equal to $1/t$. The number of links that
connect to nodes of degree $k$ at time step $t$, is equal to this
probability multiplied by the number of nodes with degree $k$,
$N p_k(t)$ and the number of links, $m$:

\[\frac{1}{t}Np_{k}\left( t \right)m = mp_{k}\left( t \right)\]

The master equation for the system then becomes:
\[
\begin{array}{ll}
(N+1)p_k(t+1) = Np_k(t) + m p_{k-1}(t) - mp_k(t) , &  k>m \\
(N+1)p_m(t+1) = Np_m(t) + 1 - mp_m(t) ,& k=m \\
p_k(t)=0 , & 0\leq k \leq m-1 \\
\end{array}
\]

The left part of the equation counts the number of nodes with degree $k$
at time $t+1$. This number is equal to the number of nodes with degree $k$
at time $t$ (first term) plus the number of nodes whose degree increases
from $k-1$ to $k$ (second term) minus the number of nodes whose degree
increases from $k$ to $k+1$ (third term).

For $t\rightarrow \infty$, the stationary state is
$p_k(t+1)=p_k(t)$, and the equations above
become:
\[
\begin{array}{ll}
p_m=\frac{1}{1+m}&\\
p_{k} = \left( \frac{m + 1}{m} \right)^{m}\frac{1}{m + 1}\left( \frac{m + 1}{m} \right)^{- k} ,
& k>m
\end{array}
\]

This confirms the well-known exponential decay with \emph{k} in random
recursive networks, where e.g. for \emph{m}=1 we have
$p_k=2^{-k}$ and for \emph{m}=2 the
distribution becomes $p_k=(3/4) 1.5^{-k}$.

\subsection*{b) Preferential selection of the initial node and \emph{q}=0}

Here we select the first node with probability proportional to its
degree, and the remaining $m-1$ nodes are selected randomly. The number of
links that point to nodes with degree $k$ are then:
\[\frac{k}{2mt} + \frac{m - 1}{t}\]
since every node has $2m$ links. The number of new links that connect to degree $k$ nodes becomes:
\[\left( \frac{m - 1}{t} + \frac{k}{2mt} \right)Np_{k}\left( t \right) = \left( m - 1 + \frac{k}{2m} \right)p_{k}\left( t \right).\]

With similar arguments as above, the master equation is then:
\[
\begin{array}{ll}
(N + 1)p_{k}( t + 1) = &\\ Np_{k}( t) + ( m - 1 + \frac{k - 1}{2m} )p_{k - 1}( t ) - ( m - 1 + \frac{k}{2m} )p_{k}( t ) & k>m \\
&\\
\left( N + 1 \right)p_{m}\left( t + 1 \right) = &\\Np_{m}\left( t \right) + 1 - \left( m - 1 + \frac{m}{2m} \right)p_{m}\left( t \right) & k=m \\
&\\
p_k(t)=0 & 0\leq k \leq m-1.\\
\end{array}
\]

The solution of the stationary state gives:
\[
\begin{array}{ll}

p_m=\frac{2}{(2m+1)} &\\
&\\
p_{k} = \frac{2}{2m + 1}\frac{A\left( A + 1 \right)\ldots(A + 2m)}{\left( A + k - m \right)\left( A + k - m + 1 \right)\ldots(A + k + m)} , &
k>m \\
\end{array}
\]
where $A=m(2m-1)$. Asymptotically, this distribution assumes the power-law
form of $k^{-(2m+1)}$, i.e. it decays much faster as we
increase \emph{m}. The dependence of the exponent on \emph{m} shows that
despite the preferential attachment rule for the first link, the network quickly behaves similar to a random structure. For m=2, for example, we get
\[p_{k} = \frac{12096}{\left( 4 + k \right)\left( 5 + k \right)(6 + k)(7 + k)(8 + k)},
k>2 \]
which behaves as k\textsuperscript{-5}.

\begin{figure}
    \centering
    \includegraphics[width=.95\linewidth]{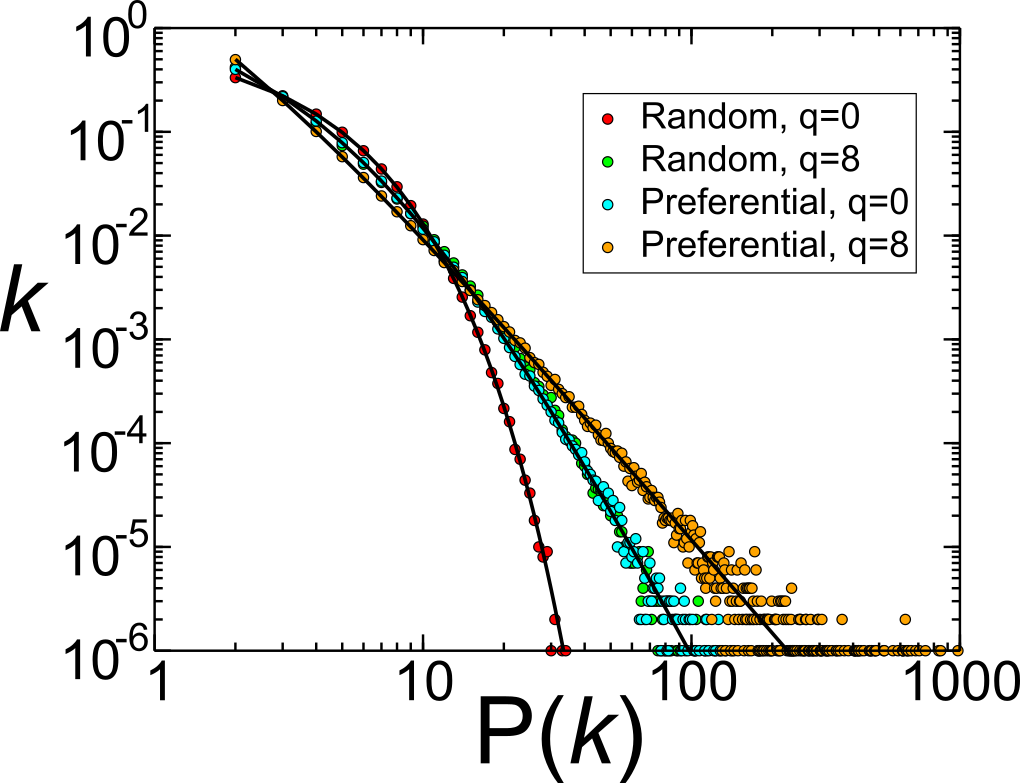}
    \caption{\textbf{Degree distributions for the propinquity model.} The
symbols represent simulations of growing networks and the lines
correspond to the exact solutions. Notice that an initial random
selection with $q=8$ gives the exact same distribution as an
initial preferential selection with $q=0$.}
    \label{figS4}
\end{figure}

\subsection*{c) Random selection of the initial node and $q\rightarrow \infty$}

For very large values of \emph{q}, the $m-1$ connections after the initial
selection almost certainly point to a neighbor of the first selected
node. The probability to select such a node then is proportional to its
degree, since this is equivalent to following the links of the first
node. This probability can be written as
\[\frac{1}{t} + \frac{k(m - 1)}{2mt}\]
and the number of new links that connect to degree $k$ nodes becomes:
\[\left( \frac{1}{t} + \frac{k\left( m - 1 \right)}{2mt} \right)Np_{k}\left( t \right) = \left( 1 + \frac{k\left( m - 1 \right)}{2m} \right)p_{k}\left( t \right).\]

With similar arguments as above, the master equation is then:

\(\left( N + 1 \right)p_{k}\left( t + 1 \right) = Np_{k}\left( t \right) + \left( 1 + \frac{(k - 1)\left( m - 1 \right)}{2m} \right)p_{k - 1}\left( t \right) - \left( 1 + \frac{k\left( m - 1 \right)}{2m} \right)p_{k}\left( t \right)\),
for $k>m$

\(\left( N + 1 \right)p_{m}\left( t + 1 \right) = Np_{m}\left( t \right) + 1 - \left( 1 + \frac{m\left( m - 1 \right)}{2m} \right)p_{m}\left( t \right)\),
for $k=m$

\(p_k(t)=0, for 0\leq k\leq m-1\).

The solution of the stationary state for m=2 has the same form as case
(b) above, i.e. when one link connects through preferential attachment
and one is selected randomly which gives a degree distribution
$k^{-5}$. We can see therefore that the
distribution changes from exponential at $q=0$ to a weak power-law
at large values of $q$.

\subsection*{d) Preferential selection of the initial node and $q\rightarrow\infty$}

This case is the same as selecting $m$ nodes preferentially, i.e. the standard Barabasi-Albert model. The known result is:

\[p_{k} = \frac{2m(m + 1)}{k(k + 1)(k + 2)}\]
which asymptotically is a power-law $k^{-3}$ with exponent
3. The effect of increasing \emph{q} leads to a broader tail in the
degree distribution, and signifies the changes in local density that
take place as we change the preferential distance of the m-1 links,
following the initial attachment.

The degree distributions for these cases are shown in Fig. S4 for $m=2$.

\section*{Fractal dimension in the propinquity model}

The concept of fractal dimension provides an efficient method for studying network structure. Typically, a network is partitioned into the smallest possible number of boxes, $N_{\rm box}$ so that within a box the maximum distance is less than $r_{\rm box}$. By varying the distance $r_{\rm box}$, we can determine if the structure has fractal features, through the exponent, $d_b$, of a possible power-law decay: $N_{\rm box} \sim r_{\rm box}^{-d_b}$. If this relation decays faster than a power-law, or equivalently for finite networks if the exponent has a large value, then the network is not fractal.

In Fig.~S5 we calculated the fractal behavior of the three empirical networks shown in Fig.~1 and we found that their structure is largely non-fractal. For example, the UC Irvine network has an approximate slope of $d_b\sim 6$ while the Facebook network does not exhibit a power law behavior at any significant range of $r_{\rm box}$.

In the case of the propinquity model, network distances remain in general small, especially for small values of $q$. This leads to a fast logarithmic drop in the number of boxes for larger values of $r_{\rm box}$ which is an indication of non-fractal behavior. At larger $q$ values there is a more prominent power-law behavior, which has a fractal dimension $d_b\sim 3$ and does not depend strongly on $q$. As a result, the fractal dimension cannot be used to distinguish networks created by the propinquity model with different $q$ values.

\begin{figure}
    \centering
    \includegraphics[width=.95\linewidth]{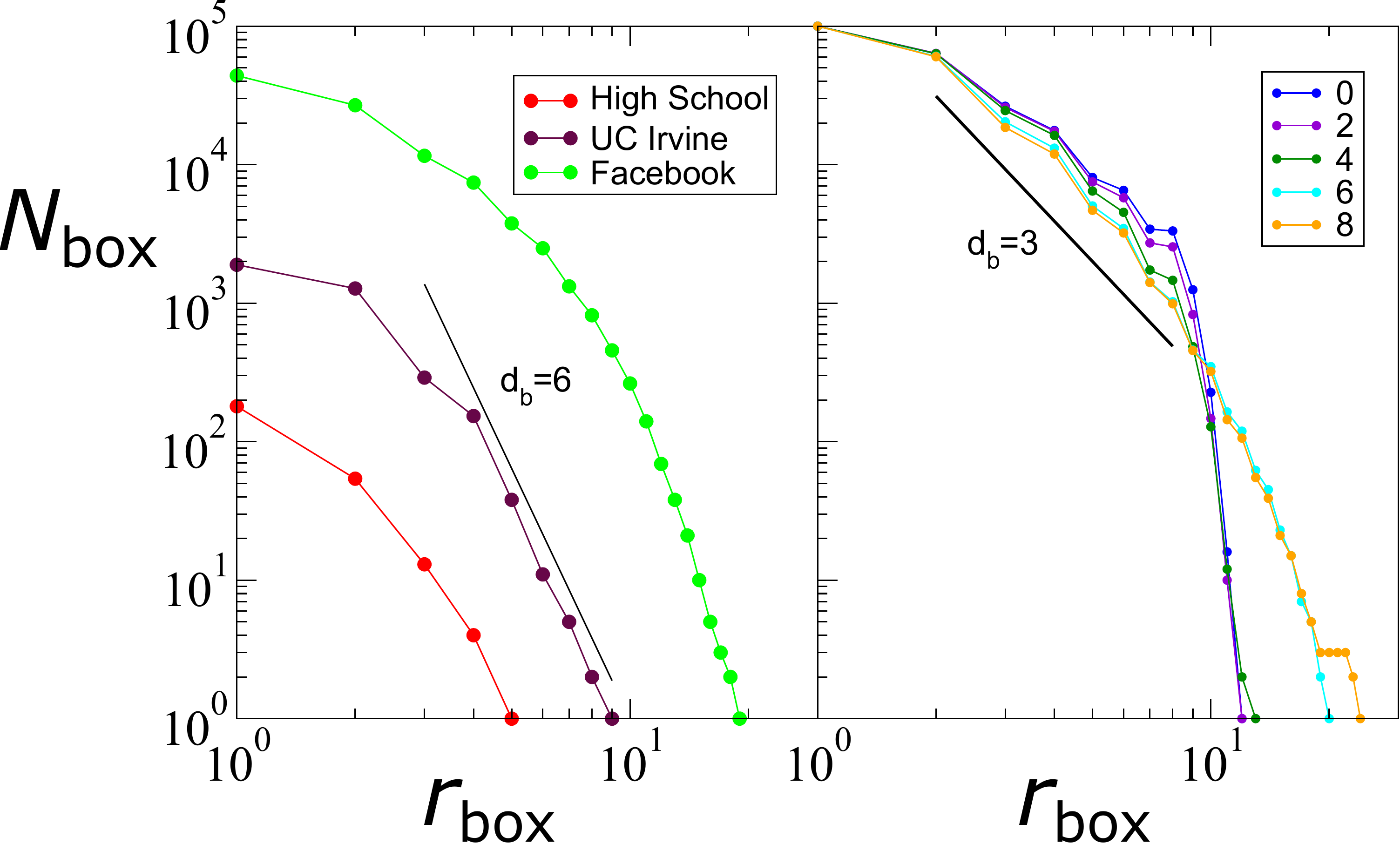}
    \caption{\textbf{Fractal behavior of the propinquity model.} The left panel shows the dependence of the minimum number of boxes $N_{\rm box}$ on the maximum distance within the box, $r_{\rm box}$ for the three networks shown in the legend. The right panel shows the same dependence for different $q$ values of the propinquity model ($q$ values are shown in the box) for networks of size $N=10^5$.}
    \label{figS5}
\end{figure}

\end{document}